# A study of aliphatic amino acids using simulated vibrational circular dichroism and Raman optical activity spectra

Aravindhan Ganesan[1*], Michael J. Brunger[2] and Feng Wang[1]

[1]eChemistry Laboratory, Faculty of Life and Social Sciences, Swinburne University of Technology, Melbourne, Victoria 3122, Australia.

[2]School of Chemical and Physical Sciences, Flinders University, GPO Box 2100, Adelaide, South Australia 5001, Australia.

*Email: aganesan@swin.edu.au

Tel: +61 3 9214 8785

## Abstract

Vibrational optical activity (VOA) spectra, such as vibrational circular dichroism (VCD) and Raman optical activity (ROA) spectra, of aliphatic amino acids are simulated using density functional theory (DFT) methods in both gas phase (neutral form) and solution (zwitterionic form), together with their respective infrared (IR) and Raman spectra of the amino acids. The DFT models, which are validated by excellent agreements with the available experimental Raman and ROA spectra of alanine in solution, are employed to study other aliphatic amino acids. The inferred (IR) intensive region (below 2000 $cm^{-1}$) reveals the signature of alkyl side chains, whereas the Raman intensive region (above 3000 $cm^{-1}$) contains the information of the functional groups in the amino acids. Furthermore, the chiral carbons of the amino acids (except for glycine) dominate the VCD and ROA spectra in the gas phase, but the methyl group vibrations produce stronger VCD and ROA signals in solution. The C-H related asymmetric vibrations dominate the VOA spectra (i.e., VCD and ROA) > 3000 $cm^{-1}$ reflecting the side chain structures of the amino acids. Finally the carboxyl and the $C_{(2)}H$ modes of aliphatic amino acids, together with the side chain vibrations, are very active in the VCD/IR and ROA/Raman spectra, which makes such the vibrational spectroscopic methods a very attractive means to study biomolecules.

## 1. Introduction

As the building blocks of life, amino acids have been one of the most investigated classes of biomolecules. Understanding their properties is for instance, important in unraveling the structures and functions of more complex biological systems such as peptides and proteins. In the gas phase (or isolated), amino acids generally exist as a neutral (NT) form, $NH_2$-CH(R)-COOH, where R is the side chain group that distinguishes between them. Study of the gas phase structures of amino acids provide insights into their intrinsic properties.[1-13] These intrinsic properties also provide references to study the molecules in their other forms, such as zwitterionic (ZW) [$NH_3^+$-CH(R)-$COO^-$], in crystal or in solution phases.

Aliphatic amino acids where the side chain consists of aliphatic alkanes, such as glycine (R-H), alanine (R-$CH_3$), valine (R-$CH(CH_3)_2$), leucine (R-$CH_2CH(CH_3)_2$) and isoleucine (R-$CH(CH_3)(CH_2CH_3)$), are important amino acids. In our previous study, we reported the ionization spectra and the electronic structures of those aliphatic amino acids.[11] Here we consider the effects of the alkyl side chain modifications on the vibrational and chiro-optical properties of these aliphatic molecules, using vibrational and vibrational optical activity (VOA) spectroscopies.

Vibrational spectroscopy is a powerful tool to reveal the structure and dynamics of biomolecules based on the molecular motions or vibrations of their structural components, such as their side chains and peptide backbones.[14] As molecular vibrations are very sensitive to inter- and intra-molecular interactions, these vibrational spectra are also useful in providing information regarding the potential binding sites among peptides, proteins and other large molecules.[14-20] For these reasons, IR and Raman spectroscopy have become important techniques to study amino acids in both the gas and aqueous phases.[1, 2, 6, 7, 14, 17, 20-30] For example, simulated IR spectra are employed to study different conformers of a number of amino acids in the gas phase,[1, 2] and also for deprotonated amino acids.[28] Very recently, Raman spectra of amino acids have been used as a reference to analyze the spectra of collagen protein within 600-1700cm$^{-1}$.[20]

Vibrational optical activity (VOA) spectroscopy is very useful to study stereo-chemical properties of bio-molecules. It measures the differential responses of a molecule to the left and right circularly polarized radiations during a vibrational transition, such as the optical and conformational characteristics of chiral molecules.[31, 32] There are two principal forms of VOA,[32, 33] vibrational circular dichroism (VCD) --- the IR sensitive form --- and Raman optical activity (ROA)---the Raman sensitive form. Both these techniques complement each other in studying the stereo-chemical properties of bio-molecules, and their principles of operation are already available in the literature.[6, 31-35] The VOA techniques are extremely useful in identifying the most favorable stereoisomers of chiral drugs, since different stereoisomers may trigger different levels of therapeutic activities against the target. A combined experimental ROA & VCD spectral study elucidated the conformers of a chiral anti-angiogenic drug, Aeroplysinin-1,[36] for example. The VCD and ROA based VOA spectroscopies have emerged as reliable tools to study the stereo-optical and conformational properties of biomolecules. There have also been significant advances in the simulations of the VOA spectroscopy, using the first principles methods. For example, the recently simulated ROA spectra of L-alanine and L-proline achieved excellent agreement with the relevant experiments, after applying anharmonic corrections such as VSCF, PT2 and VCI.[24]

Aliphatic amino acids are chiral in nature (except glycine), hence they became the target of a number of experimental and theoretical studies of the VCD and ROA spectral techniques.[5, 17, 22, 24, 29-31, 34, 37-41] As the smallest chiral amino acid, L-alanine, has been one of the most intensely studied molecules using VOA techniques for several decades.[5, 17, 22, 24, 25, 29, 30, 40] One of those studies indicated that the COOH bending vibrations are relatively intense in the VCD spectra of neutral alanine in gas phase.[17] On the other hand, in solution, the main ROA spectral features of zwitterionic (ZW) alanine are very similar in different pH conditions.[22] The ROA spectra of the alanine ZW form and its isotopomers in the $H_2O$ and $D_2O$ media reveal that the stretching and rocking modes of the methyl moiety and the $COO^-$ moiety are responsible for generating the intense ROA peaks.[30] Moreover, Yu et al found that the dominance of the methyl signals in the ROA spectrum of alanine reduces significantly when the methyl hydrogen atoms are replaced by deuterium atoms.[30] As a result, the vibrations related to $C_\alpha H$ deformation and the $COO^-$ moiety stereo-chemical correlations are responsible for the ROA signatures of the amino acids.[22, 25, 31, 37]

Fewer studies concerning the IR/VCD and Raman/ROA spectra of the amino acids and their behaviors in the gas phase and in solution have been reported, although a recent study focuses on the IR/VCD spectra of amino acids in the gas phase only.[17] As a result, we present a comprehensive *ab initio* study to simulate the VCD/IR and ROA/Raman spectra for the aliphatic amino acids. Our specific aim is to study the vibrational and chiro-optical properties of the amino acids in response to their side chains in the gas phase, and how large those changes persist when these amino acids are in solution.

## 2. Computational details

The optimized structures of the aliphatic amino acids (gas phase, NT) are identical to those from our previous study.[11] The amino acids are chiral (except for glycine) and the naturally occurring L-enantiomers of alanine, valine, leucine and isoleucine are used. The aliphatic amino acid ZW structures are optimized using the CPCM ('conductor-like polarizable continuum model')[42] water solvent model. The IR/VCD and Raman/ROA spectral simulations are performed in both the gas and solvent phases, while the VCD rotational strengths are calculated from Stephen's equation.[43, 44] All theoretical calculations are carried out using density functional theory (DFT) based on the hybrid Becke three-parameter Lee-Yang-Parr (B3LYP)[45, 46] functional combined with the 6-31++G** basis set. The B3LYP model offers the most cost-effective choice for the prediction of molecular vibrational properties, as indicated by a recent study on vibrational spectroscopy[47] and our previous studies.[8-12] It is also known that B3LYP calculated vibrational wavenumbers and VCD/ROA intensities of alanine are in good agreement with a number of experiments.[17, 30] The Gaussian 09 (G09) computational chemistry program[48] is used for our calculations, with the GaussView 5 visualization program being used to produce the art works.

## 3. Results and discussions

### 3.1 Optimized geometric parameters

Schematic descriptions of the neutral (NT) and zwitterionic (ZW) [$NH_3^+$-CH(R)-$COO^-$] forms of the amino acids are given in Fig. S1 (Supporting Information). The present study is based on the most stable NT conformers of the aliphatic amino acids, as obtained by previous studies.[1-4, 8, 49-55] Note that the structures of these NT conformers are given in the Fig. S2 (Supporting Information). In aqueous solution the proton from the carboxyl group is transferred to the amino group, thus forming $COO^-$ and $NH_3^+$ ions in the ZW forms of the amino acids. As a result, geometries of the ZW amino acids are optimized in the CPCM model. The numbering of the optimized ZW amino acids is the same as those in NT structures.

Our previous study[11] achieved excellent agreement with the available experiments, in relation to some of the properties of the NT amino acids. As a result, the present study largely concentrates on the ZW forms of the amino acids. The optimized ZW geometries of glycine and alanine with available experimental[56] and other theoretical[5, 57] results from the literature are compared in the Table S1 (Supporting Information). The geometric parameters of ZW glycine and alanine, obtained in the present study, are in good agreement with the available experimental bond lengths and bond angles, except at certain angles. For example, the maximum discrepancy in the bond lengths does not exceed 0.02 Å, whereas the discrepancy between the calculation and experiment for the $\angle C_{(1)}C_{(2)}N$ in glycine ZW is 4.6°. It is well known that the crystal structures in the experiments are "rigid", rather than the flexible structures seen in solutions or in gas phase. In addition, quantum mechanically, energies of a molecule are in general less sensitive to variations in the bond angles.[9, 11]

Table 1 compares the geometrical parameters of the NT and ZW forms of the aliphatic amino acids. The general differences between the NT and the ZW forms are seen in the local region formed by the $-COO^{-}$ and $NH_3^{+}$ moieties. This local region involves the $O_{(1)}$, $O_{(2)}$, $C_{(1)}$, $C_{(2)}$ and N atoms (refer to Fig. S2 (Supporting Information)), in which the related bond lengths, such as $C_{(1)}$-$C_{(2)}$ and $C_{(2)}$-N, increase from the NT form to the ZW form as we go from glycine to isoleucine, by approximately +0.05 Å. It is noted that the $C_{(1)}{=}O_{(2)}$ double bond lengths of the amino acids, however, stretch from 1.21 Å in the NT form to ca. 1.27 Å in the ZW form. The most significant change is that the (H)$O_{(1)}$–$C_{(1)}$ single bond lengths of 1.36 Å of the amino acids in the NT form reduce to 1.25~1.27 Å in the ZW form, a significant reduction of –0.11 Å. Moreover the (H)$O_{(1)}$–$C_{(1)}$ and $C_{(1)}{=}O_{(2)}$ bonds connect at $C_{(1)}$, indicating that the $HO_{(1)}$-$C_{(1)}{=}O_{(2)}$ moiety in the NT form takes a more delocalized network of $^{-}O_{(1)}\underline{\cdots}C_{(1)}\underline{\cdots}O_{(2)}$ in the ZW form. Other bonds, as we progress through the amino acids, such as the C-C bonds, show little changes.

The bond angles from our computations experience apparent changes from the NT to ZW form in these amino acids, in agreement with previous studies.[5, 23, 57] Similar to the bond lengths, the bond angles which change most significantly are those involved in the local region. For example, the local glycine-ZW moiety related bond angles, such as $\angle O_{(1)}C_{(1)}O_{(2)}$, $\angle O_{(1)}C_{(1)}C_{(2)}$, $\angle O_{(2)}C_{(1)}C_{(2)}$ and $\angle C_{(1)}C_{(2)}N$, all change in the same direction in going from their NT forms to the ZW counterparts. That is, if such an angle increases in going from its NT to ZW form in glycine, this angle also increases in all the other aliphatic amino acids in Table 1. For example, the $\angle O_{(1)}C_{(1)}O_{(2)}$ bond angle of the ZW form is always larger than its NT counterpart, due to loss of the H atom on the hydroxyl group $HO_{(1)}$ which forms a possible hydrogen bond with $O_{(2)}$ in the NT form. The dihedral angles further show that the four atoms, $O_{(1)}$, $C_{(1)}$, $O_{(2)}$, and $C_{(2)}$, are almost coplanar, as a result, the three angles are related as: $360°-\angle O_{(1)}C_{(1)}O_{(2)} \approx \angle O_{(1)}C_{(1)}C_{(2)}+\angle O_{(2)}C_{(1)}C_{(2)}$, in both the NT and ZW forms of the amino acids in Table 1. The angle $\angle C_{(1)}C_{(2)}N$, between the NT and ZW forms of the amino acids, reduces by approximately 8°, depending on the aliphatic side chains involved. Other bond angles outside of the local glycine-ZW moiety region exhibit only small changes of less than 3°. The most significant geometric changes for the NT and ZW forms of the amino acids are between some of the dihedral angles, which do not follow the above discussed trends in bond lengths and bond angles, indicating that the aliphatic alkyl chains play a more important role in the NT and ZW forms here.

Table 2 presents selected molecular properties, such as dipole moments, electronic spatial extents ($<R^2>$) and rotational constants, of the aliphatic amino acids in their NT and ZW forms. Available experimental and other theoretical results [25, 54, 55, 58, 59] are also given in this table for comparison purposes. It is the dipole moments of the amino acids that change most significantly, being approximately 10 fold larger in the ZW form compared to the NT form of the same compound. For example the dipole moment of NT glycine is 1.25 Debye, which becomes 13.37 Debye in the ZW form in solution. Such a significant change in dipole moment agrees with results from a recent *ab-initio* based molecular dynamics simulation.[60] This effect is due to the change in the chemical bond natures --- from covalent bonds in the NT structures to ionic bonds in the ZW forms. It is interesting that in the smaller molecules, such as glycine and alanine, the molecular size, i.e., electronic spatial extent $<R^2>$ (in a.u.), of the amino acids shrinks in going from the NT form to the ZW form, but expands in valine, leucine and isolucine. For instance, the $<R^2>$ of glycine reduces from 424.23 a.u. in the NT form to 410.23 a.u. in the ZW form, whereas the size of isoleucine increases from 1387.77 a.u. in NT to 1418.87 a.u. in ZW. This is likely to be due to the branched alkyl side chains possessed by valine, leucine and isoleucine.

### 3.2 Simulated and measured Raman/ROA spectra of ZW alanine

Fig. 1 compares the simulated and experimental[24] Raman (a) and ROA (b) spectra of ZW alanine in aqueous solution. No scaling is applied in those simulated spectra. Excellent agreement between the calculated spectra and the measured spectra is observed. The Raman spectrum of ZW alanine in the present simulation reproduces almost all the major functional spectral peaks of the measured spectra, except for the relative intensities of some of the peaks. The calculated ROA spectrum in the bottom panel of Fig. 1 also reproduces very well the experimental ROA spectrum of alanine ZW in solution, after allowing for a global shift of 38 $cm^{-1}$. Indeed, all the positive and negative bands in the calculated and experimental spectra are in excellent correlation. As a result, the present model is applied to simulate the vibrational and VOA spectra of the other aliphatic amino acids, in both gas phase (NT) and solution (ZW).

The comparison of the calculated IR and Raman spectra of the amino acids in their NT form (gas phase) is given in Fig. S3 (Supporting Information). The IR spectra are clearly dominant in the lower wavenumber region (IR active), below 2000 $cm^{-1}$, whereas the Raman spectra are more active in the higher wavenumber region (Raman active) above 3000 $cm^{-1}$. Thus the combined IR and Raman spectra are able to deliver a comprehensive picture of their behavior in this wavenumber region. Furthermore, we note that the entire region of 400-4000 $cm^{-1}$ in all the aliphatic amino acids can be divided as an alkyl-region for $\upsilon < 1600$ $cm^{-1}$ and a functional region for $\upsilon > 1600$ $cm^{-1}$. The former is mostly characterized by vibrations of the alkyl side chain in the amino acids, while the latter contains vibrations from different functional groups which are useful to differentiate the amino acids from other bio-molecules.[14, 17, 20, 21] As a result, combined IR/VCD and Raman/ROA spectra are used to assist in the study and interpretation of the vibrational & VOA (400 $cm^{-1}$ - 4000 $cm^{-1}$) spectra of the aliphatic amino acids in their NT and ZW forms.

### 3.3 Vibrational (IR and Raman) spectra of glycine

Fig. 2 presents the IR and Raman spectra of glycine in its NT form (gas phase) (a) and its ZW form (aqueous solution) (b). As the smallest amino acid, glycine does not have an α-carbon. This absence of chirality does not, therefore, produce any vibrational optical activity (i.e., VOA) signals, so that glycine does not have any VCD and ROA bands. Note that the IR intensive region (i.e., alkyl-region) and Raman intensive region (i.e., functional region), in the spectra of NT and ZW glycine, are separated by the dashed line in Fig. 2. The functional group region $\upsilon > 1600$ $cm^{-1}$, in the spectra of NT glycine, possesses five major Raman modes and two major IR modes. Those modes are assigned as the OH stretch (str) ('1'); $NH_2$ asym (asymmetric) str ('2i') and sym (symmetric) str ('2ii'); $C_{(2)}H$ asym str ('3i') and sym str ('3ii'), together with a couple of intense IR peaks due to the C=O str, OH bend ('4g'); and $NH_2$ scissoring (scis) ('5'), as marked on the spectra. It is worthy of note that the intensities of both the IR and Raman spectra in the ZW form are significantly enhanced, including a few IR spectral peaks in the high wavenumber region of 3000 $cm^{-1}$. It is also noted that the spectra of ZW glycine are correlated to the spectra of NT glycine in the gas phase. However they are not identical, that is, not all the spectral peaks in the ZW form are the same as in the NT form of glycine due to their structural differences. For example, the ZW spectral peak ‘4z’ at ~ 1680 cm$^{-1}$ is assigned to the $O_{(1)}\cdots C_{(1)}\cdots O_{(2)}$ str and $NH_3$ bending motions, whereas this peak corresponds to the one at ~1834 cm$^{-1}$ in the NT form (‘4g’). This ‘4g’ peak is in fact dominated by the C=O stretch and OH bending vibrations. In addition, the ZW spectra of glycine, compared to its NT form, does not show any peak at ~3808 cm$^{-1}$ (‘1’) due to the lack of an OH group. However, it gains an extra intensive peak (in both IR and Raman), marked as ‘2iii’ (~3500 cm$^{-1}$), which is attributed to the $NH_{(3)}$ str.

Table 3 assigns the major vibrational wavenumbers of glycine in the NT and ZW forms, as are also marked in Fig. 2, and compares them with the observed wavenumbers.[1, 27, 61] Note that both the scaled (with a scaling factor[62] of 0.96) and unscaled wavenumbers from our calculations are provided in that table. In the functional spectral region, $\upsilon$ > 1600 cm$^{-1}$, the scaled calculated wavenumbers are in good agreement with those from the experiments. For instance, the $-NH_2$ asym str vibration of NT glycine is 3470 cm$^{-1}$, which agrees well with the measured wavenumbers[1, 27] of 3410 cm$^{-1}$ and 3414 cm$^{-1}$. However it is noted that the scaling factor works less impressively in the IR intensive region at lower wavenumbers, that is, the alkyl region. On the other hand, the unscaled wavenumbers in the alkyl region are themselves sufficiently close to the measured values. For instance the $NH_2/CH_2$ wagging mode in ZW glycine is 70 cm$^{-1}$ lower than the measured value[61] when scaled, however, the unscaled value is only 17 cm$^{-1}$ away. Therefore, no scaling is applied to the calculated spectra in this study.

The alkyl region of NT glycine presents six major spectral peaks successively labeled as *a-f*, at ~1456 cm$^{-1}$ (‘a’), 1126 cm$^{-1}$ (‘b’), 919 cm$^{-1}$ (‘c’), 822 cm$^{-1}$ (‘d’), 629 cm$^{-1}$ (‘e’) and 474 cm$^{-1}$ (‘f’). These peaks are very different from those in ZW glycine that appear at 1484 cm$^{-1}$ (‘A’), 1432 cm$^{-1}$ (‘B’), 1372 cm$^{-1}$ (‘C’), 1101 cm$^{-1}$ (‘D’) and 864 cm$^{-1}$ (‘E’). Please also refer to Table 3 for their respective wavenumber assignments. The bands in the alkyl region can be relatively more complex, due to the mixed vibrational motions arising from the strong inter- and intra-molecular interactions in this species, as a consequence we therefore find differences in the NT and ZW spectra of glycine.

### 3.4 Vibrational and VOA spectra of the NT and ZW chiral amino acids

Apart from glycine, all the other aliphatic amino acids, with the bulky methyl dominant side chains, are chiral in nature with the lowest point group symmetry of $C_1$. As a result, these amino acids are optically active with VOA signals (i.e, VCD and ROA bands). In addition, due to the low point group symmetry for the amino acids, the irreducible representatives are all symmetric "a" and as a result, application of the point group symmetry information does not play a significant role in these cases. In this section, the IR based VCD spectra and Raman dependent ROA spectra are quantum mechanically simulated for both the NT and ZW forms. This was undertaken in order to explore the side chain caused structural impacts on the chiro-optical properties of those compounds. The VCD and ROA spectra, combined with their parent IR and Raman spectra respectively, are given in Figs. 3-6, whereas the assignments of the major spectral peaks are presented in tables, Table 4 and Table S2 (Supporting Information).

### 3.4.1 Alanine

Fig. 3 reports the comparative IR/VCD and Raman/ROA spectra of NT alanine (a) and ZW alanine (b). When a hydrogen atom in glycine is replaced by a methyl group, the resultant alanine molecule with an $\alpha$-carbon ($C_\alpha$) becomes chiral. As a result, optical activity signals are produced in the VCD and ROA spectra of alanine. The major spectral changes with respect to NT glycine, are that the methyl group in NT alanine lead to some additional peaks in the alkyl-region of $\upsilon < 1600$ cm$^{-1}$ and the functional region of $\upsilon > 1600$ cm$^{-1}$. Fig. 3 suggests that the amino and carboxyl vibrational motions in the functional region display very weak VCD and ROA intensities. However, the CH str vibration modes in the vicinity of 3000 cm$^{-1}$ produce a negative ROA peak at 3129 cm$^{-1}$ and a positive ROA peak at 3079 cm$^{-1}$. Those features are due to $C_{(3)}H_2$ asym str and $C_{(2)}H$ str motions respectively, as shown in Fig. 3(a) and Table S2 (Supporting information).

The VCD and ROA spectra of ZW alanine show enhanced signals. The most noticeable and intense IR spectral peak at 3016 cm$^{-1}$ appears in all the spectra of alanine in Fig. 3(b), which is assigned to the $NH_{(3)}$ vibrations (2iii). This peak, which is unique to the ZW forms of the amino acids and is not seen in the NT spectra, produces a negative peak in the VCD and the ROA spectra in ZW alanine (Fig. 3(b)). The $NH_2$ sym str (2ii) at 3489 cm$^{-1}$ corresponds to a weak positive VCD signal, whereas the $NH_2$ asym str (2i) at 3557 cm$^{-1}$ corresponds to a weak negative ROA signal. The CH vibrations clearly produce intense Raman peaks in the ZW form. For example, the peaks at 3155 cm$^{-1}$ ('3i'), 3130 cm$^{-1}$ ('3ii') and 3115 cm$^{-1}$('3iii') are mostly dominated by $C_{(2)}H$ str vibrations, although the peaks '3ii' and '3iii' also include motions of $C_{(3)}H_3$ asym str and sym str respectively. The C-H vibrations in the '3ii' peak give rise to a strong positive ROA peak and a weak negative

VCD peak at 3130 $cm^{-1}$ of ZW alanine, whereas the '3i' and '3iii' Raman peaks relate to less intense negative ROA peaks. The vibration at 1676 $cm^{-1}$ (i.e., the peak '4z') is assigned to the $COO^-$ bend and $NH_3$ deformation modes, which results in a less intense negative VCD band.

The alkyl regions of the VCD and ROA spectra ($\upsilon$ < 1600 $cm^{-1}$) exhibit more features than their parent IR and Raman spectra. Indeed some relatively intense VCD and ROA bands of ZW alanine are observed in the region of 800 $cm^{-1}$ to 1600 $cm^{-1}$. Fig. 3 and Table 4 summarize the spectral peak positions and their assignments for this molecule. The four notable peaks in the VCD spectrum of NT alanine in this region are the three negative VCD peaks at 1365 $cm^{-1}$ ('a'), 1176 $cm^{-1}$ ('b'), 1086 $cm^{-1}$ ('d') and one positive VCD band at 1131 $cm^{-1}$ ('c'). Although those peaks are produced by different vibrations, they are dominated by the $C_{(2)}H$ bending motions as assigned in Table 4. The intense bands in the VCD & ROA spectra of ZW alanine include two positive peaks at 1376 $cm^{-1}$ ('A') and 992 $cm^{-1}$ ('E'), and four negative peaks at 1358 $cm^{-1}$ ('B'), 1115 $cm^{-1}$ ('C'), 1001 $cm^{-1}$ ('D') and 878 $cm^{-1}$ ('F') in the VCD spectrum. Their corresponding ROA bands at these positions, however, switch signs. The strongest positive VCD peak at 1376 $cm^{-1}$ (A) is an intense negative peak in the ROA spectra, for example. However, a positive band at 992 $cm^{-1}$ (i.e., peak 'E') in the VCD spectrum does not appear in the corresponding ROA spectrum. As the signals here are weak, it is difficult for us to provide further analysis on this aspect. A previous experiment[24] identifies a positive ROA peak at 922 $cm^{-1}$ and a negative ROA peak at 1003 $cm^{-1}$ of ZW alanine, those are in good agreement with our calculated peak positions at 878 $cm^{-1}$ ('F') and 1003 $cm^{-1}$ ('D') respectively.

### 3.4.2 Amino acids with branched chains

Other amino acids such as valine, leucine and isoleucine are associated with more complex branched alkyl chains. For example, the terminal side chains of valine and leucine show a similar structure with a [-$CH(CH_3)_2$] group. In addition, valine and leucine both possess a single chiral carbon, i.e. $C_{(2)}$, which connects with a hydrogen, a carboxyl, an amino and an alkyl group. On the other hand, leucine and isoleucine, are isomers with similar structural variants. However isoleucine now has two chiral carbons, namely $C_{(2)}$ and $C_{(3)}$. The chirality of these compounds is the source for the VCD and ROA signals.

The IR/VCD and Raman/ROA spectra of NT and ZW valine, leucine and isoleucine are given in Figs. 4-6, accordingly. Similar to the spectra of alanine, the amino (refer to peaks labeled '2i', '2ii', '5') and carboxyl ('1' and '4g') vibrations in the functional region ($\upsilon$ > 1600 $cm^{-1}$) of NT valine, leucine and isoleucine do not produce intense VCD or ROA signals. The exception to this is for the VCD bands of NT valine (Fig. 4(a)), which exhibit intense peaks denoted as '2i', '4g' and '5'. In contrast to their NT forms, the VCD and ROA signals of the ZW amino acids in this region are more intense. For instance, the '2i' and '2ii' peaks from the sym and asym str vibrations of the amino group present a negative-positive (-ve/+ve) couplet in ZW valine, leucine and isoleucine. Due to the carboxyl $COO^{-}$ stretch of the ZW structures, the '4z' peak at 1678 $cm^{-1}$ for valine (1677 $cm^{-1}$ for leucine and 1678 $cm^{-1}$ for isoleucine) is a strong negative VCD band. It is indicative that the intra-molecular proton transfer in a ZW amino acid makes the carboxyl and amino network more optically active.

The CH stretch vibration bands, which concentrate in the wavenumber region of 3000 - 3250 $cm^{-1}$, dominate the VOA spectra of all these amino acids in both their NT and ZW configurations. It is found that the most intense ROA and VCD bands of NT valine and leucine are dominated by their chiral carbon, $C_{(2)}$, which forms a negative-positive (-ve/+ve) couplet appearing at 3065 $cm^{-1}$ (-ve)/3051 $cm^{-1}$ (+ve) in valine ('3iv'/'3v', Fig. 4(a)), and 3069 $cm^{-1}$ (-ve)/3054 $cm^{-1}$ (+ve) in leucine ('3v'/'3vi', Fig. 5(a)). Interestingly, with two chiral carbon centers, the negative ROA and VCD signals in isoleucine, which are caused by the vibrations of the chiral carbons, are apparently weaker (marked with a circle in Fig. 6(a)) than both valine and leucine. Asym str modes of the methyl groups also contribute to the signals in this wavenumber region. For example, there is an intense positive-negative couplet in the ROA spectra of valine and leucine --- 3133 $cm^{-1}$ (-ve)/3115 $cm^{-1}$ (+ve) in valine ('3i'/'3ii') and 3102 $cm^{-1}$ (-ve)/ 3108 $cm^{-1}$ (+ve) in leucine ('3iii'/'3ii'). Asym methyl vibrations in isoleucine, however, produce a positive-negative-positive triplet at 3135 $cm^{-1}$ ('3i'), 3125 $cm^{-1}$ ('3ii') and 3112 $cm^{-1}$ ('3iii'). In addition, it is noted that the signals of the VCD and ROA spectra display opposite signs. For example, an intense negative ROA peak ('3ii') at 3125 $cm^{-1}$ in NT isoleucine becomes a strongly positive peak in its VCD spectrum (see Fig. 6(a)).

The methyl vibrations dominate the VOA spectra of the ZW amino acids in solution, while the chiral carbons make important contributions to the VOA spectra of their NT counterparts. For this reason the VCD/ROA spectra of ZW valine and isoleucine share more similarities than the spectra of ZW leucine, as shown in Figs 4(b)-6(b). The most intense positive ROA band in ZW valine at 3113 $cm^{-1}$ ('3iii' in Fig. 4(b)) is assigned to the $C_{(2)}H$ str vibration. This band ('3iii') is accompanied by two methyl vibration dominant negative peaks, '3i' and '3iv' at 3121 $cm^{-1}$ and 3102 $cm^{-1}$ respectively. Similar methyl dominant spectral patterns can be found in ZW isoleucine at 3113 $cm^{-1}$ ('3ii') and 3069 $cm^{-1}$ ('3iv'). However, only one strong negative peak ('3iv') at 3096 $cm^{-1}$ is found in the ROA of leucine which is due to the $C_{(6)}H_3$ vibration. It is also noted that the intense VOA bands (i.e., VCD and ROA) are produced by the CH asym str, whereas the CH sym str do not produce strong VCD or ROA signals for both the NT and ZW amino acids. As a result, asymmetric stretching vibrations of the methyl groups dominate the functional region in the VOA spectra of the amino acids.

The spectral fingerprint or alkyl region ($\upsilon < 1600$ $cm^{-1}$) of these amino acids is side chain dependent. The assignment of the intense spectral peaks (marked in Figs 4-6), for both VCD and ROA spectra in this region, is given in Table 4. The Table S3 (Supporting Information) and Fig. S4 (Supporting Information) compare some of the vibrational wavenumbers of L-valine and the ROA spectrum of L-isoleucine, respectively, in the present work with those of the measurements reported by Gargaro et al,[25] which show reasonably good agreements. In the VCD and ROA spectra of NT valine, the most intense negative band in the VCD is at 984 $cm^{-1}$ (-ve, 'd') and that is assigned to $C_{(2)}H$ bend, $C_{(3)}H$ bend and methyl rocking modes. A couple of intense positive bands in this spectra at 969 $cm^{-1}$ ('e') and 647 $cm^{-1}$ ('f'), are assigned to the methyl rocking, OH bending, CO stretch and HNCC torsion modes. On the other hand, the most intense peak in the ROA spectra of NT valine is a couplet at 1342 $cm^{-1}$ (+ve, 'a')/1276 $cm^{-1}$ (-ve, 'b'), which are due to contributions from $C_{(2)}H$ bending, $NH_2$ twisting, $C_{(3)}H$ bending and OH bending motions. Two positive-negative spectral pairs appear in the VCD and ROA spectra of ZW valine. One such pair at 1391 $cm^{-1}$ (+ve, 'A')/1364 $cm^{-1}$ (-ve, 'B') from the $C_{(2)}H$, COO and CC stretch motions as shown in Fig. 4(b) and Table 4, is consistent with the experimental ROA bands at 1414 $cm^{-1}$ and 1357 $cm^{-1}$ in valine.[25] Another such pair is the peaks at 1144 $cm^{-1}$ ('C')/1107 $cm^{-1}$ ('D'), which are dominated by the $NH_3$ and $CH_3$ rocking modes, where the 'C' and 'D' bands swap signs between the VCD and ROA spectra.

The intense VCD peaks in the alkyl region of NT leucine and isoleucine are dominated by negative spectral peaks, as shown in Fig. 5(a) and Fig. 6(a), whereas the VCD spectra of their ZW counterparts contain more positive peaks (see Fig. 5(b) and Fig. 6(b)). The most intense VCD peak of NT leucine in this region, shown in Fig. 5(a), is a negative band ('a') at 1356 $cm^{-1}$. It is associated to combinations of OH bending, $C_{(2)}H$ stretch, CO stretch, HNCH torsion and the CC stretch motions. The other intense peaks 'b' and 'c' are also negative bands. Similarly, the VCD spectrum of NT isoleucine produces 3 negative peaks, 1343 $cm^{-1}$ ('b'), 1132 $cm^{-1}$ ('c') and 1087 $cm^{-1}$ ('d), along with a weak positive band 'a' at 1426 $cm^{-1}$. The VCD of ZW leucine displays a number of spectral peaks, which are marked as 'A', 'B', 'C', 'D' and 'E' in Fig. 5(b), in which the most intense band is a positive-negative pair at 1367 $cm^{-1}$ (+ve, 'A') and 1353 $cm^{-1}$ (-ve, 'B'). The next most intense band is also a negative-positive couplet marked as 'C' (-ve) and 'D' (+ve) that are mostly due to $NH_3$ rocking. Detailed assignments are also provided in Table 4. These VCD spectral peaks of ZW leucine disappear in the corresponding ROA spectrum, except for one strong negative ROA peak marked as 'A' at 1367 $cm^{-1}$ that is due to the bending motions of the $C_{(4)}H$, $C_{(3)}H$ and $C_{(2)}H$ groups. However, the VCD spectrum of ZW isoleucine produces the maximum number of intense bands. This includes three positive peaks ('A', 'B' and 'G') and four negative bands ('C'-'F'), whereas the ROA spectrum of ZW isoleucine only exhibits a pair of positive-negative ('C'/'D') peaks.

## 4. Conclusions

Vibrational spectra of the aliphatic amino acids are obtained in their neutral and zwitterionic forms, using density functional theory methods. We find that the calculated Raman and ROA spectrum of alanine, in aqueous solution, agrees well with the available experiment. Our study generally confirms that the functional region of 1600 - 4000 $cm^{-1}$ is more Raman active, whereas the fingerprint region of 400 - 1600 $cm^{-1}$ is IR intensive, in both the neutral (gas phase) and zwitterionic (aqueous solution) forms. In addition we further reveal that the CH dependent vibrations in the region of 3000-3250 $cm^{-1}$ dominate their VCD and ROA spectra. The chiral carbons of the neutral amino acids produce intense VCD and ROA bands, whereas the alkyl (or methyl) vibrations are more intense in the VCD and ROA signals of the zwitterionic amino acids. The relative strengths of the asymmetric and symmetric CH vibrations of the aliphatic amino acids can be indicated by the intensities of their VOA bands (i.e., VCD and ROA). The former gives intense VCD and ROA bands, while the latter exhibits weak bands in their VCD and ROA spectra. Moreover, the CH vibrations in the functional region produce opposite signs in the VCD and ROA spectra of the NT and ZW aliphatic amino acids, while the amino and carboxyl groups show identical VCD and ROA signs.

The alkyl region ($\upsilon < 1600$ $cm^{-1}$) of the VCD and ROA spectra of the amino acids are, however, dominated by the α-carbon and alkyl side chain vibrations. As a result, a combination of the vibrational (IR and Raman) and VOA (VCD and ROA) spectra can be a very useful tool in order to study comprehensively the vibrational and chiro-optical properties of amino acids and other biomolecules.

**Acknowledgements**

FW acknowledges her Vice-Chancellor's research award at Swinburne University of Technology, which supports the PhD scholarship of AG. AG thanks the ARC Centre of Excellence for Antimatter-Matter studies, Flinders University node, Australia for additional financial support. The authors would like to thank Professor Petr Bour, RNDr. Josef Kapitan and Professor Laurence D. Barron for providing their experimental Raman and ROA data, of ZW alanine, for testing the validity of our simulations. Supercomputing facilities at the National Computational Infrastructure (NCI), VPAC and the Swinburne University Supercomputer are also acknowledged.

**Table 1**: Selected geometric parameters, from the present computations, of the aliphatic amino acids in the gas (neutral or NT) and solvent (ZW) phases.

| **Parameters** | **Glycine** | | | **Alanine** | | | **Valine** | | | **Leucine** | | | **Isoleucine** | | |
|---|---|---|---|---|---|---|---|---|---|---|---|---|---|---|---|
| | **NT** | **ZW** | **\|Δ\|** | **NT** | **ZW** | **\|Δ\|** | **NT** | **ZW** | **\|Δ\|** | **NT** | **ZW** | **\|Δ\|** | **NT** | **ZW** | **\|Δ\|** |
| $O_{(1)}$-$C_{(1)}$/Å | 1.36 | 1.25 | 0.11 | 1.36 | 1.25 | 0.11 | 1.36 | 1.25 | 0.11 | 1.36 | 1.25 | 0.11 | 1.36 | 1.25 | 0.11 |
| $O_{(2)}$-$C_{(1)}$/Å | 1.21 | 1.27 | 0.06 | 1.21 | 1.27 | 0.06 | 1.21 | 1.27 | 0.06 | 1.21 | 1.27 | 0.06 | 1.21 | 1.27 | 0.06 |
| $C_{(1)}$-$C_{(2)}$/Å | 1.52 | 1.55 | 0.03 | 1.52 | 1.56 | 0.04 | 1.52 | 1.57 | 0.05 | 1.52 | 1.57 | 0.05 | 1.52 | 1.57 | 0.05 |
| $C_{(2)}$-N/Å | 1.45 | 1.50 | 0.05 | 1.45 | 1.52 | 0.07 | 1.46 | 1.52 | 0.06 | 1.45 | 1.52 | 0.07 | 1.46 | 1.52 | 0.06 |
| $C_{(2)}$-$C_{(3)}$/Å | | | | 1.53 | 1.53 | 0 | 1.55 | 1.54 | 0.01 | 1.54 | 1.53 | 0.01 | 1.55 | 1.54 | 0.01 |
| $C_{(3)}$-$C_{(4)}$/Å | | | | | | | 1.53 | 1.54 | 0.01 | 1.53 | 1.54 | 0.01 | 1.53 | 1.55 | 0.02 |
| $C_{(4)}$-$C_{(5)}$/Å | | | | | | | | | | 1.53 | 1.54 | 0.01 | 1.53 | 1.54 | 0.01 |
| ∠$O_{(1)}$-$C_{(1)}$-$O_{(2)}$/° | 123.41 | 129.02 | 5.61 | 123.12 | 128.59 | 5.47 | 123.02 | 128.52 | 5.5 | 123.14 | 128.51 | 5.37 | 122.90 | 128.47 | 5.57 |
| ∠$O_{(1)}$-$C_{(1)}$-$C_{(2)}$/° | 110.91 | 116.03 | 5.12 | 111.41 | 116.37 | 4.96 | 111.60 | 116. 47 | 4.87 | 111.51 | 116.54 | 5.03 | 111.53 | 116.58 | 5.05 |
| ∠$C_{(1)}$-$C_{(2)}$-N/° | 115.56 | 107.32 | 8.24 | 113.67 | 105.66 | 8.01 | 113.04 | 105.41 | 7.63 | 113.60 | 105.40 | 8.2 | 113.10 | 105.31 | 7.79 |
| ∠N-$C_{(2)}$-$C_{(3)}$/° | | | | 109.64 | 111.32 | 1.68 | 111.12 | 113.08 | 1.96 | 110.16 | 111.81 | 1.65 | 111.22 | 113.29 | 2.07 |
| ∠$C_{(1)}$-$C_{(2)}$-$C_{(3)}$/° | | | | 108.40 | 113.72 | 5.32 | 109.21 | 114.30 | 5.09 | 107.64 | 113.48 | 5.84 | 109.59 | 114.72 | 5.13 |
| ∠$C_{(2)}$-$C_{(3)}$-$C_{(4)}$/° | | | | | | | 110.03 | 111.90 | 1.87 | 114.35 | 116.22 | 1.87 | 111.09 | 112.80 | 1.71 |
| ∠$C_{(3)}$-$C_{(4)}$-$C_{(5)}$/° | | | | | | | | | | 109.30 | 109.51 | 0.21 | 113.37 | 113.80 | 0.43 |
| ∠$O_{(1)}$-$C_{(1)}$-$C_{(2)}$-N/° | 180.00 | -179.94 | 0.06 | 161.62 | 178.54 | 16.92 | 165.10 | -179.63 | 14.53 | 161.30 | 179.78 | 18.48 | 168.04 | -179.10 | 11.06 |
| ∠$O_{(2)}$-$C_{(1)}$-$C_{(2)}$-N/° | 0.00 | 0.10 | 0.1 | -20.01 | -1.00 | 19.01 | -16.50 | 1.36 | 17.86 | -20.41 | 0.61 | 21.02 | -13.20 | 2.14 | 15.34 |
| ∠$O_{(1)}$-$C_{(1)}$-$C_{(2)}$-$C_{(3)}$/° | | | | -76.20 | -59.10 | 17.10 | -70.67 | -54.83 | 15.84 | -76.42 | -57.52 | 18.9 | -67.30 | -53.81 | 13.49 |
| ∠$O_{(2)}$-$C_{(1)}$-$C_{(2)}$-$C_{(3)}$/° | | | | 102.10 | 121.37 | 19.27 | 107.70 | 126.18 | 18.48 | 101.81 | 123.31 | 21.5 | 111.50 | 127.44 | 15.94 |
| ∠$C_{(1)}$-$C_{(2)}$-$C_{(3)}$-$C_{(4)}$/° | | | | | | | 169.30 | 172.74 | 3.44 | -179.76 | 173.56 | 6.2 | -65.41 | -59.74 | 5.67 |
| ∠N-$C_{(2)}$-$C_{(3)}$-$C_{(4)}$/° | | | | | | | -65.31 | 66.61 | 131.92 | -55.42 | -67.36 | 11.94 | 60.43 | 61.27 | 0.84 |
| ∠$C_{(2)}$-$C_{(3)}$-$C_{(4)}$-$C_{(5)}$/° | | | | | | | | | | -179.61 | 172.60 | 7.01 | 171.78 | 167.09 | 4.69 |

**Table 2**: Selected molecular properties, from the present computations, of the aliphatic amino acids in the gas (neutral or NT) and solvent phases (ZW). Corresponding experimental results, where available, are also shown.

| Parameters | Glycine | | Alanine | | Valine | | Leucine | | Isoleucine | |
|---|---|---|---|---|---|---|---|---|---|---|
| | NT[59]* | ZW | NT[63]* | ZW | NT[55]* | ZW | NT[58]* | ZW | NT[54]* | ZW |
| Dipole Moment (Debye) | 1.25 (1.10) | 13.37 | 1.35 (1.80) | 13.15 | 1.37 (1.48) | 12.95 | 1.10 (1.14) | 13.24 | 1.28 (1.20) | 12.92 |
| $<R^2>$ (a.u.) | 424.23 | 410.23 | 581.21 | 580.51 | 1046.61 | 1063.28 | 1572.06 | 1623.32 | 1387.77 | 1418.87 |
| **Rotational Constants** | | | | | | | | | | |
| A (GHz) | 10.26 (10.34) | 10.35 | 5.07 (5.07) | 4.84 | 2.94 (2.98) | 2.96 | 2.76 (2.75) | 2.82 | 2.10 (2.09) | 2.15 |
| B (GHz) | 3.87 (3.87) | 4.06 | 3.05 (3.10) | 3.36 | 1.44 (1.43) | 1.46 | 0.85 (0.85) | 0.83 | 1.11 (1.11) | 1.09 |
| C (GHz) | 2.90 (2.91) | 3.02 | 2.30 (2.26) | 2.20 | 1.34 (1.32) | 1.27 | 0.80 (0.79) | 0.76 | 0.98 (0.97) | 0.93 |

*Values given in parentheses are experimental values from the corresponding references cited.

**Table 3**: Present scaled and unscaled vibrational wavenumbers of NT (in gas phase) and ZW (in solution) glycine, along with the available experimental data (all wavenumbers are given in $cm^{-1}$).

| No.* | Glycine NT | | | | No* | Glycine ZW | | | |
|---|---|---|---|---|---|---|---|---|---|
| | This Work (Unscaled) | This Work (Scaled)^ | EXP[#] | Assignment | | This Work (Unscaled) | This Work (Scaled)^ | EXP[61] | Assignment |
| 1 | 3808 | 3656 | 3560 | OH str | 2(i) | 3561 | 3419 | | $NH_2$ asym str |
| 2(i) | 3615 | 3470 | 3410(3414) | $NH_2$ asym str | 2(ii) | 3496 | 3356 | | $NH_2$ sym str |
| 2(ii) | 3535 | 3394 | | $NH_2$ sym str | 3(i) | 3190 | 3062 | | $CH_2$ asym str |
| 3(i) | 3112 | 2988 | (3084) | $CH_2$ asym str | 3(ii) | 3129 | 3004 | 2960 | $CH_2$ sym str |
| 3(ii) | 3070 | 2947 | 2958(2920) | $CH_2$ sym str | 2(iii) | 3055 | 2933 | 2930 | $NH_{(3)}$ sym str |
| 4g | 1835 | 1762 | 1779(1703) | C=O str, OH bend | 4z | 1680 | 1612 | 1630 | $C_{(1)}O_{(1)}$ str, $NH_{(3)}$ bend |
| 5 | 1678 | 1611 | 1630(1610) | $NH_2$ scis | | 1665 | 1598 | 1603 | $NH_3$ deformation |
| a | 1457 | 1399 | 1429(1410) | $CH_2$ scis | 5 | 1636 | 1570 | 1566 | $NH_2$ scis |
| | 1402 | 1346 | 1373 | $CH_2$ wag, COH bend | A | 1484 | 1425 | 1430 | $CH_2$ scis |
| | 1382 | 1327 | (1334) | $CH_2$ / $NH_2$ twist | B | 1432 | 1375 | | $NH_3$ puckering |
| | 1305 | 1253 | | $CH_2$ wag, OH bend | C | 1372 | 1317 | 1403 | $CO_{(2)}$ str, CC str, $CH_2$ wag |
| | 1180 | 1133 | | $CH_2$ twist. $NH_2$ twist | | 1324 | 1271 | 1341 | $CH_2$ / $NH_2$ wag |
| | 1164 | 1117 | 1136 | CN str, $CH_2$/$NH_2$ wag, CO str | | 1301 | 1249 | 1314 | $CH_2$ / $NH_2$ twist |
| b | 1126 | 1081 | 1101 | CO str, OH bend, CN str | D | 1101 | 1057 | 1171 | $CH_2$ wag, $NH_3$ rock |
| c | 919 | 882 | 907 | $CH_2$ wag, $NH_2$ wag | | 1097 | 1053 | | $CH_2$ twist, $NH_3$ rock |
| | 914 | 877 | 883(893) | $CH_2$ rock, $NH_2$ twist | | 983 | 944 | | CN str, HCNH tor |
| d | 822 | 789 | 801 | CC str, $NH_2$ wag, $CH_2$ wag | | 929 | 892 | | $CH_2$ / $NH_2$ rock |
| | 630 | 605 | 619 | COOH tor, HNCC tor | E | 864 | 830 | 961 | CC str, $CH_2$ wag |
| e | 629 | 604 | | OH OP bend, $CH_2$ rock | | 675 | 648 | 674 | CCO bend, CCN tor |
| f | 474 | 455 | 500 | OH OP bend, $CH_2$ rock | | 570 | 548 | | $CH_2$ rock |
| | 463 | 444 | 463 | HCCN tor, CCO bend, COOH tor | | 499 | 479 | | CCO bend, HCCN tor |

*No. represents the peaks marked in Fig.. 4; #Ref. 13[13] (values given in parentheses are from Ref. 27[27]).
^Scaled by a factor of 0.96.[62]

**Table 4**: Present results showing the comparative vibrational wavenumbers (in $cm^{-1}$) and their assignments, of the NT and ZW forms of the aliphatic amino acids in the alkyl-region ($\upsilon$ < 1600 $cm^{-1}$).

| Molecules | Vibrational modes |
|---|---|
| Alanine (NT) | **(a)** 1365 - $C_{(2)}H$ bend, CO str, OH bend; **(b)**1176 -$C_{(2)}H$ bend, CN str, HNCH tor; **(c)** 1131 -$C_{(2)}H$ bend, CO str, OH bend; **(d)** 1086 -$C_{(2)}H$ bend, $C_{(3)}H_3$ rock, HCCH tor.; |
| Alanine (ZW) | **(A)** 1376 -$C_{(2)}H$ bend; **(B)** 1358 -$C_{(2)}H$ bend, $C_{(2)}$-$C_{(3)}$ str; **(C)** 1115 -$C_{(2)}H$ bend, $C_{(3)}H_3$ rock, $NH_3$ rocking; **(D)** 1001 -$C_{(2)}H$ bend, $C_{(3)}H_3$ rock, $NH_3$ rock; **(E)** 992 - $C_{(2)}H$ bend, $C_{(3)}H_3$ rock, $NH_3$ rock; **(F)** 878 - HCCH tors., $C_{(1)}C_{(2)}$ str, $C_{(2)}N$ str. |
| Valine (NT) | **(a)** 1342 – CC str, CO str, OH bend, $NH_2$ twist, $C_{(2)}H$ bend; **(b)** 1276 -$C_{(2)}H$ str, OH bend, $C_{(3)}H$ bend; **(c)** 1134 -CO str, OH bend; **(d)** 984 -$C_{(2)}H$ bend, $C_{(3)}H$ bend, $C_{(5)}H_3$, $C_{(2)}H_3$ Rocking; **(e)**969 -$C_{(4)}H_3$/$C_{(5)}H_3$ rocking, $C_{(2)}$-$C_{(3)}$ str, $C_{(2)}H$ bend; **(f)** 647 -OH bend, CO str, HNCC tors, $C_{(2)}H$ bend. |
| Valine (ZW) | **(A)** 1391 -$C_{(2)}H$ bend, COO str, $C_{(2)}H$ bend; **(B)** 1364 -COO str, CC str, $NH_3$ puckering, $C_{(2)}H$ bend; **(C)** 1144 -$NH_3$ rocking, $CH_3$ rocking, HCCH tors; **(D)** 1107 -$NH_3$ rocking, $C_{(2)}H$ bend. |
| Leucine (NT) | **(a)** 1356 -OH bend, CO str, HNCH tors, CC str, $C_{(4)}H$ bend, $C_{(2)}$ bend; **(b)** 1179 -$C_{(2)}H$ bend, OH bend, HNCH tors; **(c)** 967 -$NH_2$ wag, HCCH tors, $C_{(2)}H$ bend, CC str; **(d)** 740 -COOH tors, $C_{(2)}H$ bend, HCCC tors; |
| Leucine (ZW) | **(A)** 1367 - $C_{(4)}H$ bend, $C_{(3)}H$/$C_{(2)}H$ bend; **(B)** 1353 -$C_{(2)}H$ bend, CC str, $C_{(4)}H$ bend, $NH_3$ puckering; **(C)** 1072 -$C_{(2)}H$ bend, $NH_3$ rocking, $C_{(3)}H_2$ Rocking; **(D)** 1034 -CN str, $NH_3$ rocking, HCCH tors; **(E)** 927 - $C_{(6)}H_3$/$C_{(5)}H_3$ rocking, $C_{(4)}H$ bend. |
| Isoleucine (NT) | **(a)** 1426 - $NH_2$ twist, $C_{(2)}H$ bend; **(b)** 1343 -$NH_2$ twist, $C_{(2)}H$ bend, $C_{(3)}H$ bend, $C_{(4)}H$ bend, OH bend; **(c)** 1132 - OH bend, CO str, $C_{(2)}H$ bend; **(d)** 1087 -$CH_3$ rocking, $C_{(3)}H$ bend; |
| Isoleucine (ZW) | **(A)** 1424 -$CH_3$ rocking; **(B)** 1341-$C_{(4)}H_2$ wagging, $C_{(2)}H$ bend, CC str; **(C)** 1366 -$C_{(4)}H_2$ twist, $C_{(2)}H$ bend; **(D)** 1348 - COO str, $NH_3$ puckering, $C_{(2)}H$ bend, $C_{(3)}H$ bend; **(E)**1136 -$CH_3$ rock, HCCH tors, HNCC tors; **(F)** 1106 -$NH_3$ tors, $C_{(2)}H$ bend; **(G)** 1003 - HCCH tors, $CH_3$ rock, CC str. |

**Figure captions**

Fig. 1: Simulated and experimental[24] Raman and ROA spectra of the alanine ZW in aqueous solution.

Fig. 2: Simulated IR and Raman spectra of the (a) NT (gas phase) and (b) ZW (aqueous solution) glycine molecule. Please refer to Table 3 for the wavenumbers and vibrational modes of the peaks marked in the spectra.

Fig. 3: Simulated IR/VCD and Raman/ROA spectra of the (a) NT and (b) ZW forms of alanine. Please refer to Table 4 and Table S2 (Supporting Information) for the wavenumbers and vibrational modes marked in the spectra.

Fig. 4: Simulated IR/VCD and Raman/ROA spectra of the (a) NT and (b) ZW forms of valine in gas and aqueous solution respectively. Please refer to Table 4 and Table S2 (Supporting Information) for the wavenumbers and vibrational modes marked in the spectra.

Fig. 5: Simulated IR/VCD and Raman/ROA spectra of the (a) NT and (b) ZW forms of leucine. Again refer to Table 4 and Table S2 (Supporting Information) for the wavenumbers and vibrational modes marked in the spectra.

Fig. 6: Simulated IR/VCD and Raman/ROA spectra of the (a) NT and (b) ZW forms of isoleucine, in gas and aqueous solution respectively. Again refer to Table 4 and Table S2 (Supporting Information) for the wavenumbers and vibrational modes marked in the spectra.